\documentclass[aps,prl,reprint,groupedaddress,showpacs,amsmath,amssymb,ocolumn,floatfix,longbibliography]{revtex4-1}
\usepackage{graphicx}
\usepackage{bm}
\usepackage{color}
\usepackage[normalem]{ulem}
\usepackage{amsmath}
\usepackage{tabularx}
\usepackage{lineno}

\usepackage[colorlinks=true]{hyperref} 
\hypersetup{
    unicode=false,          % non-Latin characters 
    pdftoolbar=true,        % show Acrobat
    pdfmenubar=true,        % show Acrobat 
    pdffitwindow=false,     % window fit to page when opened
    pdfstartview={FitH},    % fits the width of the page to the window
    pdftitle={XXX},    % title
    pdfauthor={XXX et al.},     % author
    pdfsubject={},   % subject of the document
    pdfcreator={},   % creator of the document
    pdfproducer={}, % producer of the document
    pdfkeywords={} {} {}, % list of keywords
    pdfnewwindow=true,      % links in new window
    colorlinks=true,       % false: boxed links; true: colored links
    linkcolor=magenta, %red,          % color of internal links (change box color with linkbordercolor)
    citecolor=blue,        % color of links to bibliography
    filecolor=magenta,      % color of file links
    urlcolor=blue           % color of external links
} 
\usepackage{cleveref}

\newcommand{\Rpara}{$R_{\parallel}$}

\usepackage{pifont}

\usepackage{bbm}

\begin{document}

\title{A Hierarchy of Superconductivity and Topological Charge Density Wave States in Rhombohedral Graphene}

\author{Ron Q. Nguyen$^{1}$}
\thanks{These authors contributed equally to this work.}
\author{Hai-Tian Wu$^{1}$}
\thanks{These authors contributed equally to this work.}
\author{Erin Morissette$^{1}$}
\thanks{These authors contributed equally to this work.}
\author{Naiyuan J. Zhang$^{1}$}
\thanks{These authors contributed equally to this work.}
\author{Peiyu Qin$^{1}$}
\author{K. Watanabe$^{2}$}
\author{T. Taniguchi$^{3}$}
\author{Aaron W. Hui$^{1}$}
\author{D. E. Feldman$^{1,5}$}
\author{J.I.A. Li$^{1,4}$}
\email{jia.li@austin.utexas.edu}

\affiliation{$^{1}$Department of Physics, Brown University, Providence, RI 02912, USA}
\affiliation{$^{2}$Research Center for Functional Materials, National Institute for Materials Science, 1-1 Namiki, Tsukuba 305-0044, Japan}
\affiliation{$^{3}$International Center for Materials Nanoarchitectonics,
National Institute for Materials Science,  1-1 Namiki, Tsukuba 305-0044, Japan}
\affiliation{$^{4}$Department of Physics, University of Texas at Austin, Austin, TX 78712, USA}
\affiliation{$^5$Brown Theoretical Physics Center, Brown University, Providence, Rhode Island 02912, USA}

\date{\today}

\maketitle

\textbf{
Superconductivity and the quantum Hall effect are conventionally regarded as mutually exclusive: superconductivity is suppressed by magnetic fields, whereas the quantum Hall effect relies on them. Here we report a striking exception, where an unconventional superconducting phase is stabilized by an out-of-plane magnetic field and coexists with a re-entrant integer quantum Hall (RIQH) effect in moir\'e-less rhombohedral hexalayer graphene. The re-entrant quantum Hall state, arising from a bubble-like charge density wave (CDW), provides a natural backdrop for the emergence of superconductivity. Angle-resolved transport reveals that the field-stabilized superconducting phase occupies the same density--displacement-field regime as a stripe-ordered phase at zero field ~\cite{Koulakov1996stripe,Lilly1999stripe,Lilly1999stripe2,Gervais_3rdLL,Halperin2020FQHE,Sammon2019stripe,Morissette2025stripeSC}, yet only develops once the stripe is replaced by a bubble-like CDW at finite field~\cite{Fogler2002stripe,Gervais_3rdLL,Eisenstein2002bubble,Goerbig2003RIQH,Goerbig2004electronsolid,Xia2004bubble,Deng2012reentrant,Liu2012reentrant,Halperin2020FQHE,Chen2019RIQH}. These findings demonstrate a decisive role of CDW order in stabilizing superconductivity in rhombohedral graphene, establishing a new paradigm for the interplay between superconductivity and quantum Hall physics.
}

\begin{figure*}
\includegraphics[width=1\linewidth]{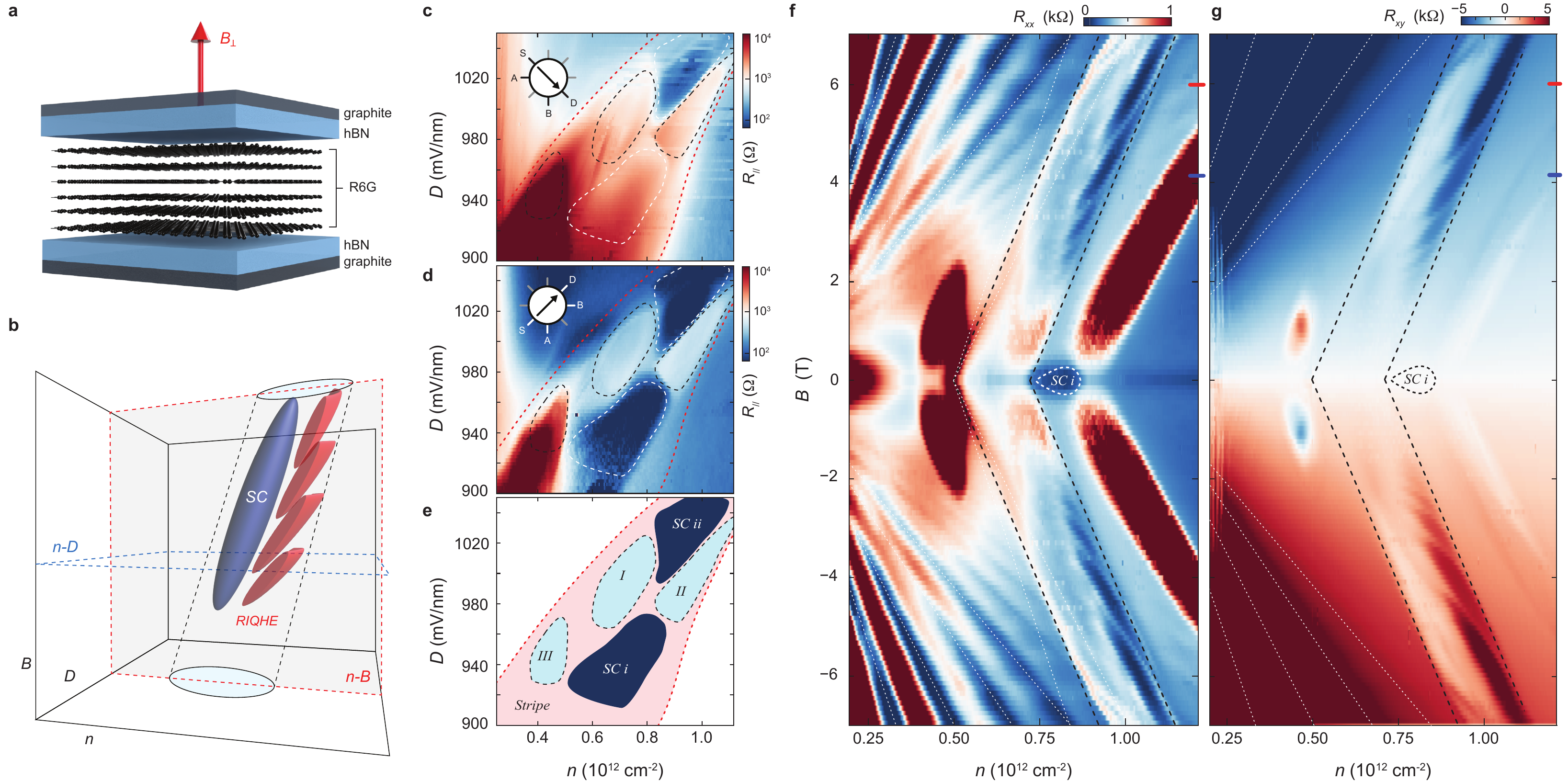}
\caption{\label{fig1} \textbf{Multi-dimensional phase space of rhombohedral hexalayer graphene.} 
(a) Schematic of the R6G heterostructure, with the red arrow indicating the out-of-plane magnetic field $B_{\perp}$.  
(b) Schematic phase diagram defined by carrier density $n$, displacement field $D$, and out-of-plane magnetic field $B_{\perp}$. 
(c, d) $n$--$D$ maps of longitudinal resistance measured at $B_{\perp} = 0$, with current flowing along the principal axes of (c) maximum and (d) minimum resistivity.  
(e) Annotated $n$--$D$ map highlighting prominent phases at $B_{\perp} = 0$, including the stripe phase and striped superconductivity.  
(f, g) $n$--$B$ maps at $D = 967$~mV/nm, showing (f) longitudinal resistance $R_{xx}$ and (g) transverse resistance $R_{xy}$. White dashed lines trace the trajectories of integer quantum Hall states, while black dashed lines mark the boundaries of regime~I. Red and blue bars denote the $B_{\perp}$ values at which the line traces in Fig.~\ref{fig2}a--d are taken.
}
\end{figure*}

\begin{figure*}
\includegraphics[width=0.99\linewidth]{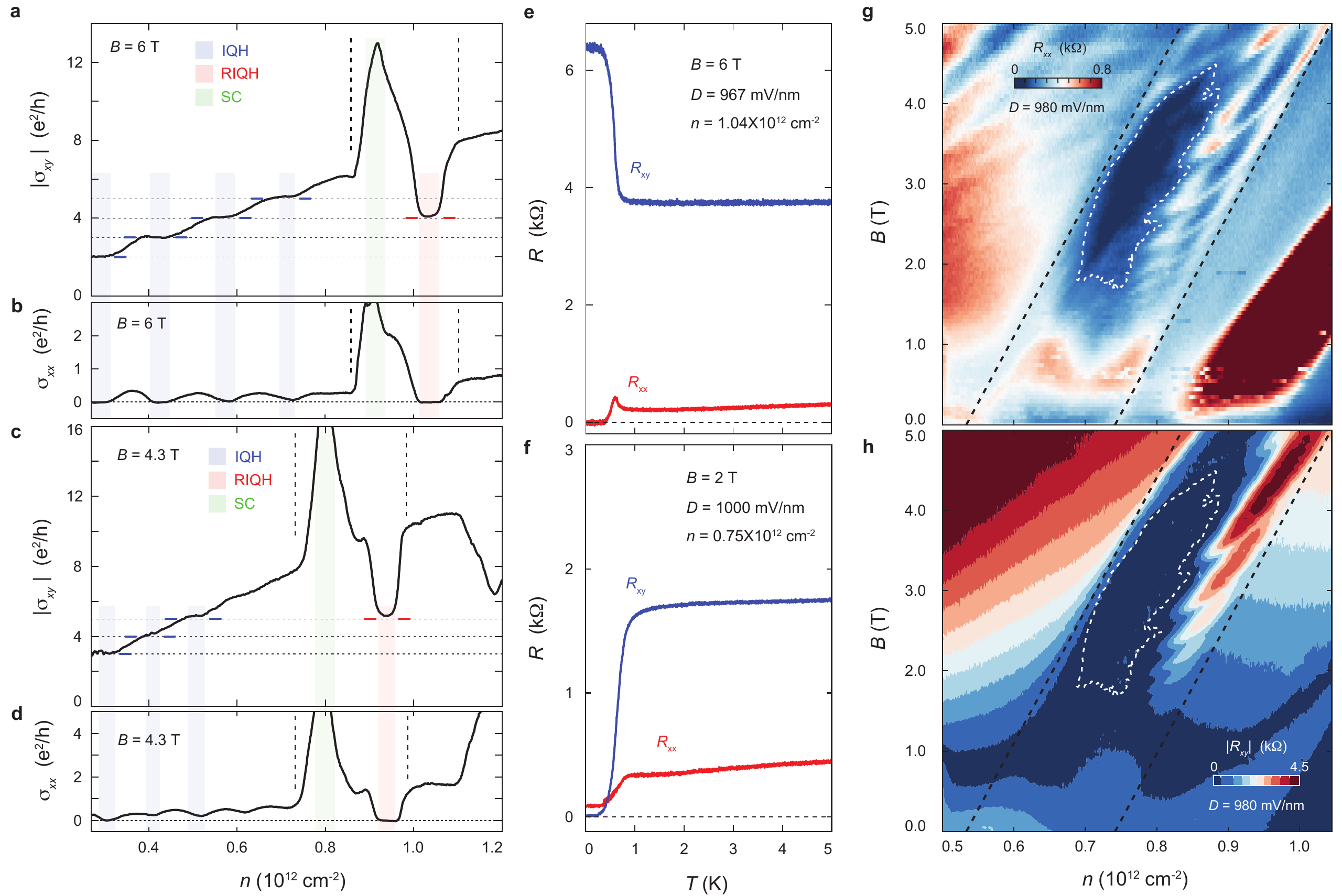}
\caption{\label{fig2} 
\textbf{RIQH states and $B_{\perp}$-stabilized superconductivity.} 
(a--d) Transport response showing IQH states, RIQH states, and the $B_{\perp}$-stabilized superconducting phase, highlighted by blue, red, and green shading, respectively. Inside regime I (indicated by two vertical dashed lines), RIQH states are identified by re-entrant plateaus in the Hall conductivity and vanishing longitudinal conductivity, whereas the superconducting phase exhibits diverging conductivity in both longitudinal and transverse channels.
(e, f) Temperature dependence of $R_{xx}$ (red) and $R_{xy}$ (blue) measured from (g) the adjacent RIQH state and (h) the $B_{\perp}$-stabilized superconducting phase.  
(g, h) $n$--$B$ maps of (g) $R_{xx}$ and (h) $R_{xy}$ at $D = 980$~mV/nm. The superconducting region is outlined by a white dashed line. The sequence of RIQH states emerges on the high-density side of the superconducting phase.
}
\end{figure*}

\begin{figure*}
\includegraphics[width=0.95\linewidth]{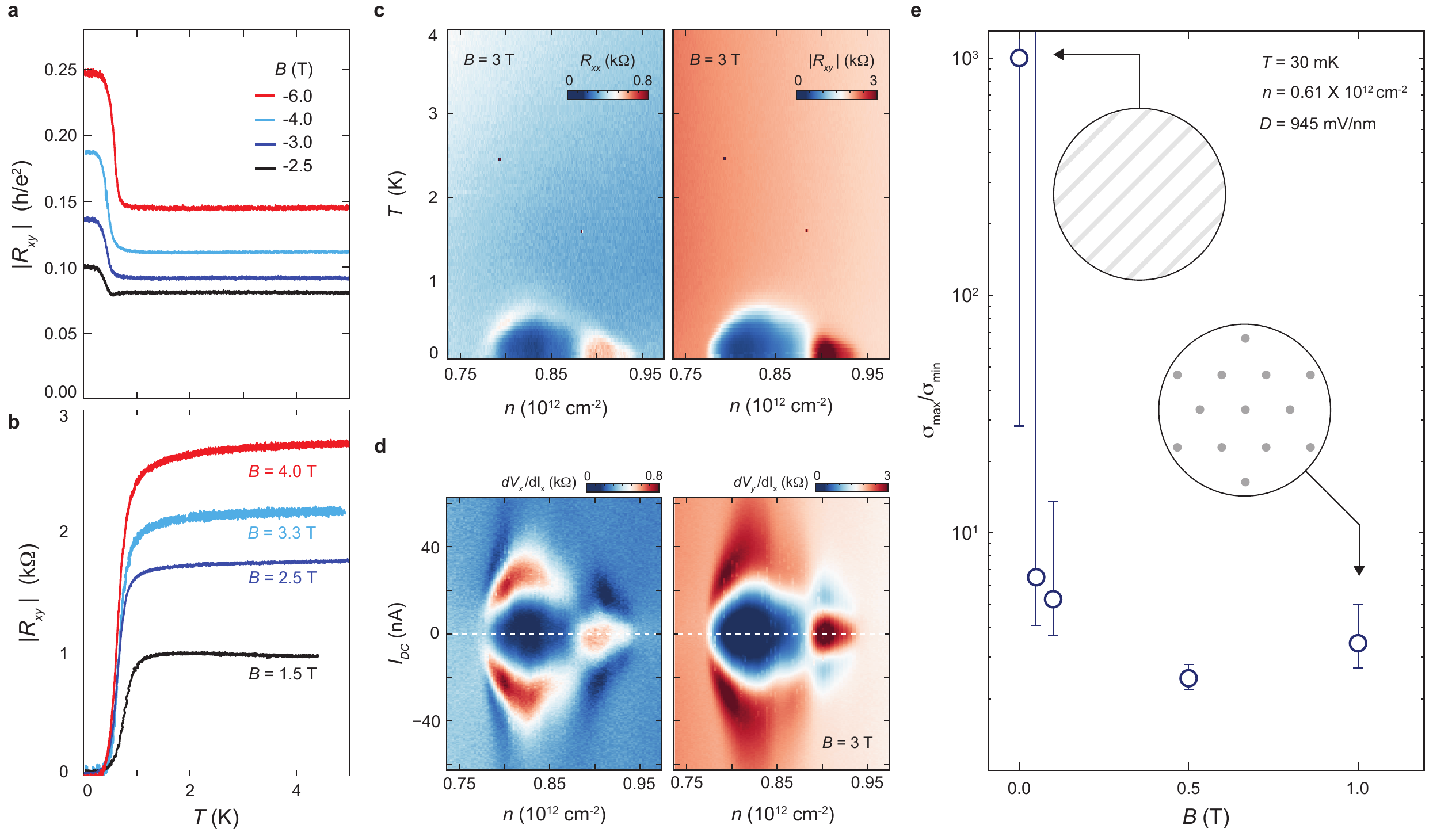}
\caption{\label{fig3} \textbf{Stripe and bubble-like phases.} 
(a, b) Temperature dependence of $|R_{xy}|$ measured at different $B_{\perp}$ values within (a) RIQH states and (b) the $B_{\perp}$-stabilized superconducting phase.  
(c) Longitudinal resistance $R_{xx}$ (left) and transverse resistance $R_{xy}$ (right) as functions of carrier density $n$ and temperature $T$ at $B_{\perp} = 3$~T and $D = 1000$~mV/nm.  
(d) Differential resistance $dV/dI$ as a function of $n$ and d.c. current bias $I_{\mathrm{DC}}$ at $B_{\perp} = 3$~T and $D = 1000$~mV/nm. Data in the left (right) panel were obtained in the longitudinal (transverse) configuration.  
(e) Transport anisotropy, defined by the conductivity ratio $\sigma_{\mathrm{max}}/\sigma_{\mathrm{min}}$, as a function of $B_{\perp}$. Insets: schematic diagrams of two distinct CDW orders—a stripe phase that produces extreme anisotropy at zero field, and a bubble-like phase stabilized at finite $B_{\perp}$.
}
\end{figure*}

\begin{figure*}
\includegraphics[width=0.9\linewidth]{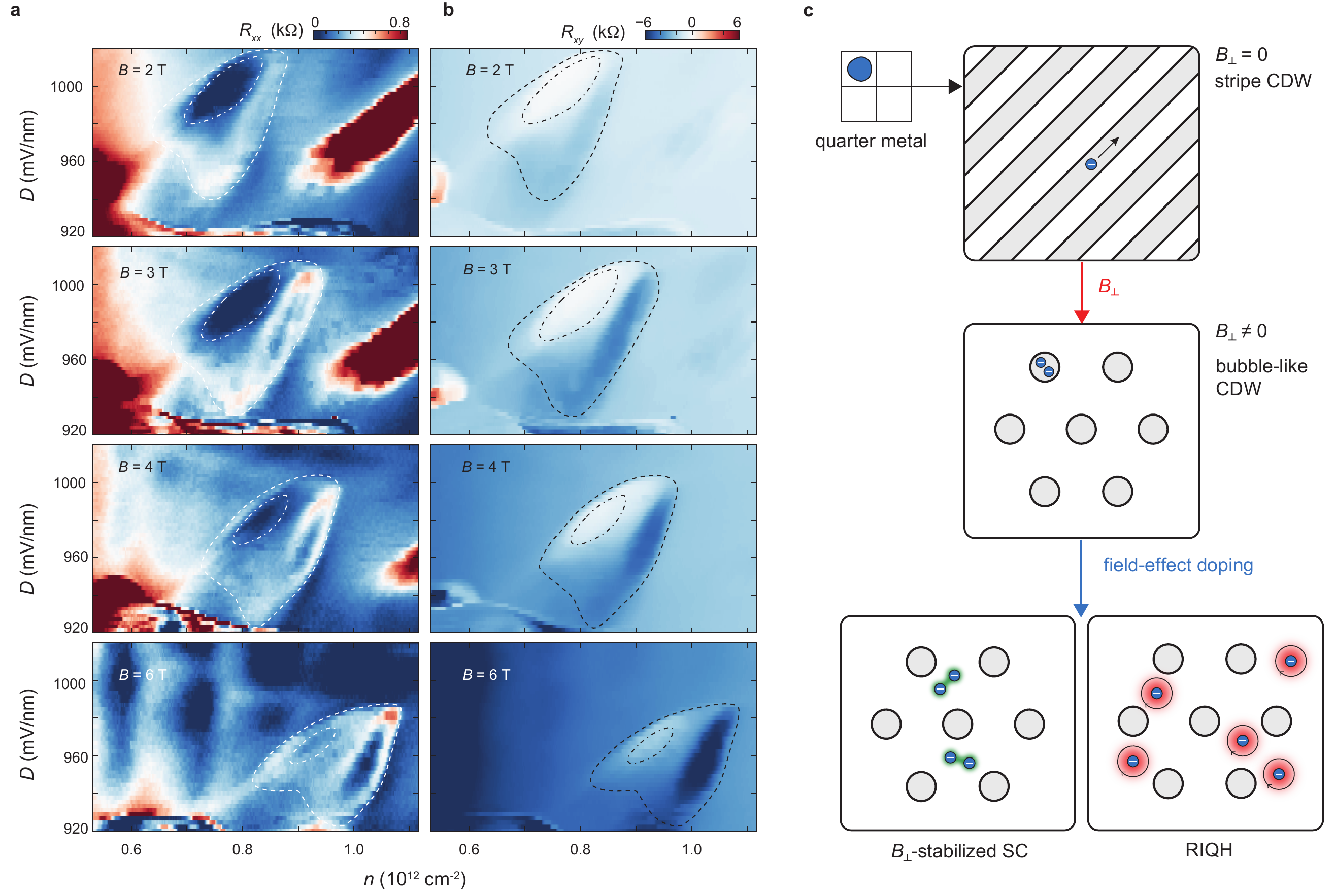}
\caption{\label{fig4}{\bf{A hierarchy between superconductivity and CDW orders.}}  
$n$--$D$ maps of (a) $R_{xx}$ and (b) $R_{xy}$ measured at different values of $B_{\perp}$. The dashed line outlines the boundary of regime I, while the dash-dotted line encloses the superconducting phase. (c) Schematic diagrams illustrating the hierarchy of topology, CDW orders, and superconductivity. In the relevant phase space, the emergence of a quarter metal state, evidenced by the anomalous Hall effect, appears at the highest temperature ~\cite{Morissette2025stripeSC}. Extreme transport anisotropy is observed upon further decreasing temperature, indicating a stripe CDW phase at $B_{\perp} = 0$. The application of $B_{\perp}$ suppresses the stripe order, stabilizing a bubble-like CDW state. Here, electrons are localized into insulating islands, shown as gray circles. The blue arrows represent the non-trivial topology associated with the bubble-like CDW order. As additional charges are added through field-effect doping, itinerant electrons occupy the regions between insulating islands, giving rise to superconductivity on the low-density side of regime I and RIQH states on the high-density side of regime I.
}
\end{figure*}

An enduring theme in the study of superconductivity is its fundamental antagonism with magnetic fields: superconducting phases are typically destroyed by even modest fields. It is therefore striking when superconductivity persists under large fields, defying conventional wisdom. Such behavior sets high-$T_c$ superconductors apart from conventional counterparts, establishing a vibrant research frontier for decades~\cite{ReviewHighTcNRP2021}. Even more exotic is the scenario in which a superconducting phase is not merely resilient to magnetic fields, but is in fact \textit{stabilized} by them—implying an unconventional pairing symmetry. A prominent example is UTe$_2$, a candidate spin-triplet superconductor that exhibits field-induced reentrant superconductivity under strong magnetic fields~\cite{Ran2019Extreme,Ran2019UTe2,Aoki2019UTe2,Lewin2023UTe2}.

In two-dimensional (2D) electron systems, a superconductor that persists under a large out-of-plane magnetic field is especially appealing, as it could enable direct interplay with the quantum Hall effect. Such interplay has long been envisioned as a pathway to non-Abelian anyons and topological superconductivity, with potential applications in fault-tolerant quantum computation~\cite{Nayak2008RMP,DasSarma2015Majorana,PeraltaGavensky2020Majorana,Alicea2016Parafermions}. Yet the inherent fragility of 2D superconductors typically precludes this coexistence. To date, progress has largely relied on engineered hybrid platforms, where superconducting and quantum Hall regions are coupled across artificial interfaces~\cite{Amet2016Supercurrent,Lee2017QHEdges,Gul2022AndreevFQHE,Zhao2023SC_QHE}. More recently, coexistence between superconductivity and zero-field analogs of quantum Hall states has been demonstrated in various 2D material heterostructures~\cite{Stepanov2020,Choi2025SC}. This includes reports of superconductivity adjacent to fractional Chern insulators near zero field, which have reignited interest in anyonic superconductivity and its topological implications ~\cite{Xu2025unconventional,Nosov2025anyonSC,Pichler2025anyonSC,ShiSenthil_FQAH_doping_2024}.

Here, we uncover a distinct form of interplay between superconductivity and the quantum Hall effect in moir\'e-less rhombohedral hexalayer graphene (R6G) ~\cite{Zhou2021RTG,Zhou2021RTG_SC,Zhou2022BLG,Han2025chiral,Choi2025SC}. In sharp contrast to all previously studied two-dimensional superconductors, where superconductivity is invariably suppressed by magnetic fields, in R6G, the superconductor is instead stabilized by an out-of-plane field, emerging in close proximity to a hierarchy of re-entrant integer quantum Hall (RIQH) states~\cite{Eisenstein2002bubble,Fogler2002stripe,Goerbig2003RIQH,Goerbig2004electronsolid,Xia2004bubble,Deng2012reentrant,Gervais_3rdLL,Liu2012reentrant,Halperin2020FQHE,Chen2019RIQH}. By examining the RIQH states in tandem with angle-resolved transport, we demonstrate an unusual hierarchy linking CDW orders and superconducting phases. At zero field, an electronic smectic (stripe) phase is manifested by extreme transport anisotropy~\cite{Koulakov1996stripe,Lilly1999stripe,Lilly1999stripe2,Gervais_3rdLL,Halperin2020FQHE,Sammon2019stripe,Morissette2025stripeSC}, coexisting with a striped superconductor that inherits this anisotropy~\cite{Morissette2025stripeSC}. Upon application of a finite perpendicular magnetic field, both the stripe order and striped superconductor are suppressed, giving way to a bubble-like CDW order, revealed through the emergence of RIQH states~\cite{Eisenstein2002bubble,Xia2004bubble,Deng2012reentrant,Gervais_3rdLL,Liu2012reentrant,Fogler2002stripe,Halperin2020FQHE}. This evolution of the underlying CDW order gives rise to the field-stabilized superconductor.

The R6G sample is assembled into a dual-encapsulated heterostructure, as illustrated in Fig.~\ref{fig1}a. This device geometry defines a low-temperature phase space governed by three independently tunable parameters: carrier density $n$, displacement field $D$, and out-of-plane magnetic field $B_{\perp}$. As schematically shown in Fig.~\ref{fig1}b, both superconductivity and the sequence of RIQH states are confined within a well-defined region of the $n$--$D$--$B$ space, whose boundary resembles a tilted cylinder. 

At zero field ($B_{\perp} = 0$), the $n$--$D$ phase space is dominated by an electronic smectic phase, previously identified through its extreme transport anisotropy~\cite{Morissette2025stripeSC}. This anisotropy is evident in Figs.~\ref{fig1}c--d, which show longitudinal resistance measured along two orthogonal directions corresponding to maximum and minimum resistivity. The pronounced discrepancy between these directions reveals a strongly anisotropic state that spans much of the $n$--$D$ map, further confirmed by angle-resolved transport measurements (see Fig.~\ref{angular}). This anisotropy has been attributed to a smectic order, highlighted by the red-shaded area in Fig.~\ref{fig1}e. Within this region, vanishing resistance signals the presence of two superconducting phases at zero field—denoted SC~i and SC~ii. Properties of SC~i were examined in a previous report, where it was shown to emerge concurrently with the smectic order and to inherit its strong directional character~\cite{Morissette2025stripeSC}.

The $n$--$D$ map at zero field also reveals several metallic pockets with transport responses distinct from their surroundings, outlined by black dashed contours in Figs.~\ref{fig1}c--e. For clarity, we label these regions~I, II, and III. Regime~I corresponds to the $B_{\perp}=0$ cross-section of the cylindrical volume illustrated in Fig.~\ref{fig1}b. In the following, we use transport measurements to characterize the electronic orders that emerge around this cylindrical phase space, while additional discussions of regimes~II and III are provided in the Methods.

In the $n$--$B$ plane at fixed displacement field $D = 967$~mV/nm, the evolution of regime I is delineated by black dashed boundaries, marked by peaks in the longitudinal resistance $R_{xx}$ and abrupt jumps in the Hall resistance $R_{xy}$ (Figs.~\ref{fig1}f--g). These boundaries enclose the portion of phase space that hosts both the $B$-stabilized superconducting phase and the sequence of RIQH states. On the high-density side of regime~I, the dashed contours further define the volume occupied by SC~i, which is suppressed by an out-of-plane field of $B_{\perp}=0.5$~T.

At a finite $B_{\perp}$, the emergence of RIQH states is demonstrated by the density dependence of $\sigma_{xx}$ and $\sigma_{xy}$, shown in Figs.~\ref{fig2}a--d and extracted from line traces of $R_{xx}$ and $R_{xy}$ in Figs.~\ref{fig1}f--g. Outside regime~I, delineated by the vertical dashed lines in Figs.~\ref{fig2}a--d, the evolution of $\sigma_{xx}$ and $\sigma_{xy}$ follows a conventional sequence of integer quantum Hall (IQH) states, forming a standard Landau fan emanating from the charge neutrality point (CNP).

In contrast, transport behavior departs from this IQH sequence within regime~I. Here, $\sigma_{xy}$ develops reentrant plateaus quantized at the same values as IQH states at lower carrier density, while $\sigma_{xx}$ simultaneously vanishes to zero. These features—highlighted by the red vertical stripes—are characteristic signatures of RIQH states ~\cite{Eisenstein2002bubble,Xia2004bubble,Deng2012reentrant,Gervais_3rdLL,Liu2012reentrant,Halperin2020FQHE,Chen2019RIQH}.

The re-entrant Hall response indicates an effective reduction of charge carrier density in R6G. As the sample shows no evidence of alignment with the hexagonal boron nitride substrate, this reduction is most naturally attributed to the formation of CDW order. Figure~\ref{fig2}e presents the temperature dependence of a representative RIQH state, revealing a sharply-defined transition at $T \approx 0.4$~K: $R_{xy}$ deviates from the re-entrant plateau while $R_{xx}$ rises from zero. This temperature-driven transition is qualitatively distinct from that of an IQH state. Rather than the gradual, monotonic thermal activation characteristic of IQH states (see Fig.~\ref{RT_QHE}b), $R_{xx}$ here displays a non-monotonic temperature dependence with a pronounced resistive peak. Such behavior is widely associated with the melting transition of an underlying CDW order, providing further support for CDW formation underlying the RIQH state.

Interestingly, on the low-density side of regime~I, both $\sigma_{xy}$ and $\sigma_{xx}$ exhibit a dramatic enhancement compared to adjacent IQH states, which we attribute to the onset of a superconducting phase. As shown in Fig.~\ref{fig2}f, the temperature dependence of the transport response, measured at nearby values of $D$ and $B_{\perp}$ where superconductivity is more robust, exhibits a pronounced transition at $T \approx 0.4$~K. Below this temperature, both $R_{xx}$ and $R_{xy}$ vanish, consistent with the emergence of a superconducting state.

The $B_{\perp}$-stabilized superconductor coexists with a sequence of RIQH states. This is revealed by plotting transport responses across the $n$--$B_{\perp}$ plane (Figs.~\ref{fig2}g--h), where boundaries of regime I are marked by black dashed lines.

On the low-density side of regime~I, the superconducting phase is identified by vanishing $R_{xx}$ and $R_{xy}$, shown in dark blue on the chosen color scale and enclosed by the white dashed line. The lower-field boundary of both the superconducting and RIQH states is marked by a pronounced resistive peak in $R_{xx}$, which appears across regime~I around $B_{\perp} \approx 1$~T (Fig.~\ref{fig2}g). The superconducting phase persists up to $B_{\perp} > 4$~T, demonstrating unprecedented field stability for a two-dimensional superconductor. This unusual dependence on $B_{\perp}$ provides strong evidence for the unconventional character of the superconducting phase. In this high-field regime, the superconductivity likely arise from spin-triplet pairing in an odd orbital angular momentum channel, consistent with the possibility of $p$-wave pairing symmetry~\cite{Anderson1961}.  

On the high-density side, a sequence of RIQH states is characterized by vanishing $R_{xx}$ together with re-entrant Hall plateaus. Notably, these RIQH states follow distinct trajectories across the $n$--$B_{\perp}$ map, each associated with a different re-entrant Hall plateau. For example, the RIQH state at $B_{\perp} = 6$~T (Figs.~\ref{fig2}a--b) exhibits a plateau at $\sigma_{xy} = 4e^2/h$, whereas another at $B_{\perp} = 4.3$~T (Figs.~\ref{fig2}c--d) displays a plateau at $\sigma_{xy} = 5e^2/h$. Collectively, these RIQH states form a hierarchy that closely resembles the Landau fan of IQH states. However, the slopes of their trajectories deviate from those expected for the corresponding quantized Hall plateaus (Fig.~\ref{mismatch}), a discrepancy naturally explained by the presence of an underlying CDW order~\cite{Huang2025apparent}.

Conventionally, RIQH states emerge in the quantum Hall regime when 2D electrons are tuned to specific Landau level (LL) filling fractions ~\cite{Eisenstein2002bubble,Xia2004bubble,Deng2012reentrant,Gervais_3rdLL,Liu2012reentrant,Halperin2020FQHE,Chen2019RIQH}. The formation of a CDW order freezes the partially filled LL, causing the Hall response to reenter the quantized value of the nearest integer plateau. In contrast, the emergence of RIQH in R6G induces the Hall response to reenter into a distant integer plateau. In the absence of a moir\'e superlattice, this behavior represents a qualitatively new mechanism of CDW instability, which is decoupled from LL formation.

In the following, we discuss the connection between the RIQH and superconducting states. 

First, the two phases appear to share a common origin, supported by several lines of experimental evidence. As shown in Fig.~\ref{fig3}a, $R$--$T$ curves measured from different RIQH states all display a sharply defined onset at $T \approx 0.4$~K. This onset temperature remains essentially constant over a wide range of $B_{\perp}$, indicating that the underlying CDW order is largely insensitive to Landau level formation. This suggests that the CDW order is not confined to the RIQH states, but instead extends across a broader portion of the surrounding phase space. Consistently, the temperature dependence of superconducting transport (Fig.~\ref{fig3}b) reveals a transition temperature that closely tracks the melting temperature of the CDW order. A natural interpretation is that the CDW order remains stable throughout regime~I, serving as the parent order from which both the RIQH and superconducting phases emerge.

The is further supported by the transport response across regime~I (Figs.~\ref{fig3}c--d). An $n$--$T$ map at fixed $B_{\perp}$ and $D$ (Fig.~\ref{fig3}c) shows that the superconducting transition and the onset of RIQH states trace out a continuous envelope that evolves smoothly with carrier density $n$. A similar correspondence is observed in the current–voltage (I--V) characteristics (Fig.~\ref{fig3}d). In the superconducting phase, a sufficiently large d.c.~bias drives a superconductor-to-metal transition, defining the critical supercurrent. Strikingly, a d.c.~current of comparable magnitude also destabilizes the RIQH state, causing the Hall response to deviate from its re-entrant plateau. These observations show that the underlying CDW order is fragile against both elevated temperature and d.c.~bias, and that this fragility shapes the boundaries of the superconducting and RIQH states in Figs.~\ref{fig3}c--d. Taken together, these coordinated behaviors provide further confirmation that superconductivity and RIQH states emerge from the same underlying CDW instability. 

Secondly, superconductivity and RIQH states also compete with one another, most clearly at the boundary of the superconducting phase. Along the trajectory of a RIQH state, the superconducting region is suppressed, producing an indentation in its phase boundary. Conversely, away from RIQH trajectories, the superconducting region expands slightly. The variation in the superconducting boundary highlights the mutual competition between the two phases.

Since the zero-field phase space is dominated by a smectic order, we now turn to the relationship between this stripe phase and the CDW order underlying the RIQH states. To this end, we examine the evolution of transport anisotropy, quantified by the ratio of maximum to minimum conductivity, $\sigma_{\max}/\sigma_{\min}$, extracted from angle-resolved transport measurements~\cite{Chichinadze2024nonlinearHall}. As shown in Fig.~\ref{fig3}e, the smectic order at $B_{\perp}=0$ exhibits extreme anisotropy, with $\sigma_{\max}/\sigma_{\min} \approx 10^3$. This ratio decreases rapidly with increasing $B_{\perp}$, reaching $\sigma_{\max}/\sigma_{\min} \approx 2$ at $B_{\perp}=0.5$~T. The strong suppression of anisotropy demonstrates that the CDW order at finite $B_{\perp}$ is qualitatively distinct from the zero-field smectic phase and is instead consistent with a bubble-like phase~\cite{Fogler2002stripe,Xia2004bubble,Eisenstein2002bubble,Liu2012reentrant,Gervais_corbino_reentrant}. This bubble-like phase likely exhibits spatial distortions that weakly break rotational symmetry, but it does not display the extreme anisotropy characteristic of stripe order.

Two key observations suggest that the relevant CDW orders are topologically nontrivial ~\cite{Tan2024hallcrystal,Dong2024hallcrystal,Soejima2024hallcrystal,Dong2024AHC,Patri2024QAH,Dong2024RPG,Lu_2025}, in stark contrast to conventional settings where charge density wave (CDW) phases arise from non-topological electronic bands. First, transport in regime~I displays pronounced hysteresis loops as $B_{\perp}$ is swept back and forth (Fig.~\ref{Bloop}), a hallmark of nontrivial band topology. This hysteresis is consistent with previous reports in R4G and R5G over similar $n$--$D$ regimes, where an anomalous Hall effect arises from a quarter-metal state~\cite{Han2025chiral}. Second, the phase boundaries of regime~I, marked by white and black dashed lines in Fig.~\ref{fig4}a--b (and by black dashed lines in Fig.~\ref{fig1}g--h and Fig.~\ref{fig2}g--h), shift significantly in both $n$ and $D$ with varying $B_{\perp}$; similar field-dependent shifts are also observed for regimes~II and III (Fig.~\ref{fig1}f--g and Fig.~\ref{Pockets}). Such $B_{\perp}$-dependence points to a nontrivial topological character of the underlying CDW order. Notably, in the $n$--$B$ plane, the boundaries of regime~I trace a fractional Streda slope of $t = 2.5$, typically associated with the fractional quantum Hall effect at filling fraction $\nu = 5/2$ ~\cite{Streda1983streda,Willet1987,Zibrov2017,Li.17b}. Although the absence of a fully developed incompressible state precludes a definitive identification, the emergence of a fractional slope raises an important open question concerning the topological nature of the CDW order in this system.

Altogether, these observations reveal a hierarchical organization between topology, CDW orders, and superconductivity, as illustrated in Fig.~\ref{fig4}c. In the relevant phase space of R6G, strong Coulomb interactions generate a sequence of emergent phases underpinned by nontrivial topology. At the primary level of this hierarchy, a quarter-metal state appears at the highest temperature, manifested through the anomalous Hall effect (Fig.~\ref{Bloop})~\cite{Morissette2025stripeSC}. This quarter-metal phase likely provides the nontrivial topology inherited by the subsequent orders. At zero field, the simultaneous observation of extreme transport anisotropy and an anomalous Hall effect identifies a stripe CDW phase with nontrivial topology~\cite{Morissette2025stripeSC}. Upon application of an out-of-plane magnetic field, this stripe order evolves into what is likely a bubble-like CDW phase, evidenced by the suppression of transport anisotropy and the emergence of RIQH states. The bubble-like order also carries nontrivial topology, as indicated by the fractional Středa slope. At the highest level of the hierarchy, competing superconducting and re-entrant quantum Hall phases emerge as itinerant electrons are introduced into the CDW background.

These itinerant electrons play a central role in enabling both superconductivity and the RIQH states. On the low-density side of regime~I, a pairing instability among itinerant electrons gives rise to the observed superconducting phase (Fig.~\ref{fig4}b). We note that this scenario is distinct from a pair-density-wave (PDW) order~\cite{Agterberg2020PDW}. In a PDW, superconductivity is the primary order and the density wave emerges as a secondary modulation, whereas in the present case the situation is reversed: a primary CDW order hosts a secondary superconducting phase. On the high-density side, by contrast, the out-of-plane magnetic field quantizes the itinerant electrons into Landau levels, producing an effective IQHE. Because the underlying CDW reduces the effective density of itinerant carriers, the resulting Hall plateaus align with those expected for IQHE states emanating from the CNP, giving rise to the observed re-entrant behavior.

As temperature increases, the CDW order melts, destroying the spatial segregation between localized and itinerant carriers. This melting transition naturally accounts for the sharply defined thermal onset observed in both the superconducting and quantum Hall responses (Fig.~\ref{fig2}e--f and Fig.~\ref{fig3}a--b). Indeed, the superconducting phase, the sequence of RIQH states, and the boundaries of regime~I all vanish once the temperature is raised slightly above the CDW melting transition, as illustrated in Fig.~\ref{R_n_temp} and Fig.~\ref{PnBT}. 

Another remarkable observation is that the structure of the CDW varies continuously as $n$, $D$, and $B_{\perp}$ are tuned across the phase space. For instance, the fraction of localized carriers within the CDW order can be tuned by the out-of-plane displacement field $D$. As shown in Figs.~\ref{fig4}a--b (see also Fig.~\ref{5T}), the re-entrant plateaus trace a sloped trajectory in the $n$--$D$ plane. At fixed $B_{\perp}$, each plateau corresponds to a constant density of itinerant electrons. The finite slope therefore indicates that, at fixed $n$, the relative fractions of itinerant and localized carriers shift with varying $D$. This provides direct evidence that the structure of the CDW order is tunable by the displacement field. 

A complementary signature comes from the apparent mismatch between the quantized Hall plateaus and the slopes of the RIQH states across the $n$--$B$ plane (Fig.~\ref{mismatch}). This deviation points to a continuous evolution of the CDW structure with magnetic field. In other words, in the presence of a malleable CDW order, the Streda slope becomes ill-defined for RIQH states~\cite{Huang2025apparent}.

Lastly, we comment on what appears to be a universal link between superconducting phases and their associated CDW orders. Beyond the results presented here, where the coexistence with RIQH states demonstrates a direct connection between unconventional superconductivity and a bubble-like CDW order, a similar correspondence is found for the zero-field superconducting phase, \textit{SC i}. Previous work has shown that \textit{SC i} inherits the extreme transport anisotropy characteristic of the surrounding stripe order~\cite{Morissette2025stripeSC}. The stability of \textit{SC i} closely mirrors that of the stripe phase: not only do \textit{SC i} and the stripe order share a similar onset temperature, but both are also suppressed at comparable values of $B_{\perp}$. Furthermore, the coexistence of unconventional superconductivity and neighboring RIQH states echoes a recent report in twisted transition metal dichalcogenides~\cite{Xu2025unconventional}. Together, these findings suggest that unconventional superconductivity is generically intertwined with coexisting CDW orders, pointing to a unifying principle for stabilizing unconventional superconductivity in two-dimensional electron systems. This framework provides a new lens for identifying, understanding, and ultimately engineering exotic superconducting states in strongly interacting quantum materials.

\section*{acknowledgments}

J.I.A.L. wishes to acknowledge helpful discussions with Andrea Young. This material is based on the work supported by the Air Force Office of Scientific Research under award no. FA9550-23-1-0482.
R.Q.N., N.J.Z, and J.I.A.L. acknowledge support from the Air Force Office of Scientific Research. E.M. acknowledge support from U.S. National Science Foundation under Award DMR-2143384.  K.W. and T.T. acknowledge support from the JSPS KAKENHI (Grant Numbers 21H05233 and 23H02052) and World Premier International Research Center Initiative (WPI), MEXT, Japan. 
Part of this work was enabled by the use of pyscan (github.com/sandialabs/pyscan), scientific measurement software made available by the Center for Integrated Nanotechnologies, an Office of Science User Facility operated for the U.S. Department of Energy. D.E.F. was supported in part by the National Science Foundation under grant No. DMR-2204635.

\bibliography{Li_ref}

\newpage
\clearpage

\section*{Methods}

\renewcommand{\thefigure}{M\arabic{figure}}

\def\theequation{M\arabic{equation}} 
\def\thetable{M\Roman{table}}
\setcounter{figure}{0}
\setcounter{equation}{0}

In this section, we provide detailed discussions to further substantiate the results presented in the main text. Our focus is on transport signatures of the RIQH states, angle-resolved transport response of the stripe and bubble-like phase.  Additionally, we elaborate on the extraction of intrinsic anisotropy from the conductivity matrix and discuss the subtleties of interpreting angle-resolved transport data in the presence of stripe order, particularly under conditions of extreme anisotropy.

\subsection{Transport signatures of RIQH states}

\begin{figure*}
\includegraphics[width=0.95\linewidth]{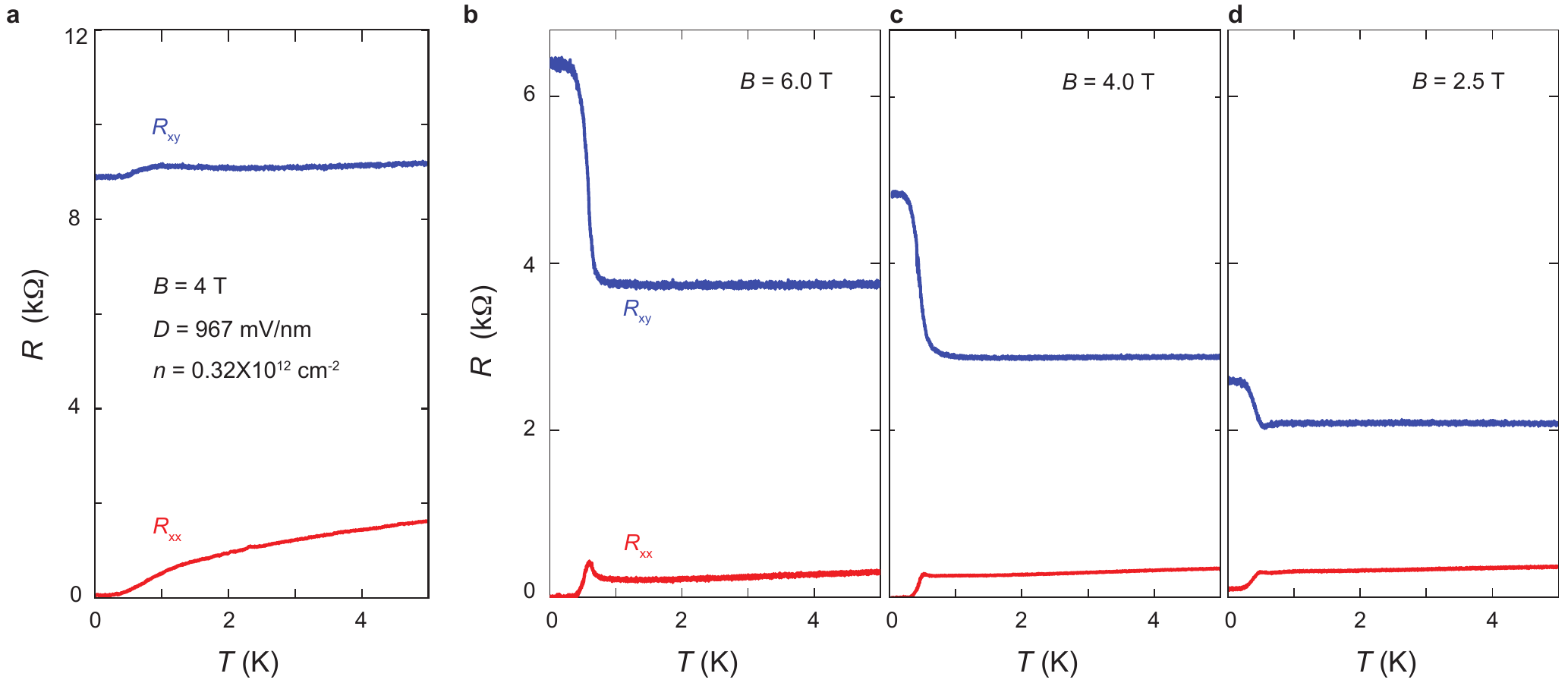}
\caption{\label{RT_QHE} \textbf{Transport signatures of quantum Hall states.} 
Temperature dependence of longitudinal resistance $R_{xx}$ (red traces) and Hall resistance $R_{xy}$ (blue traces): 
(a) in an IQHE state emerging from the charge neutrality point, and 
(b) in RIQH states inside regime~I at different values of $B_{\perp}$.  
}
\end{figure*}

\begin{figure*}
\includegraphics[width=0.92\linewidth]{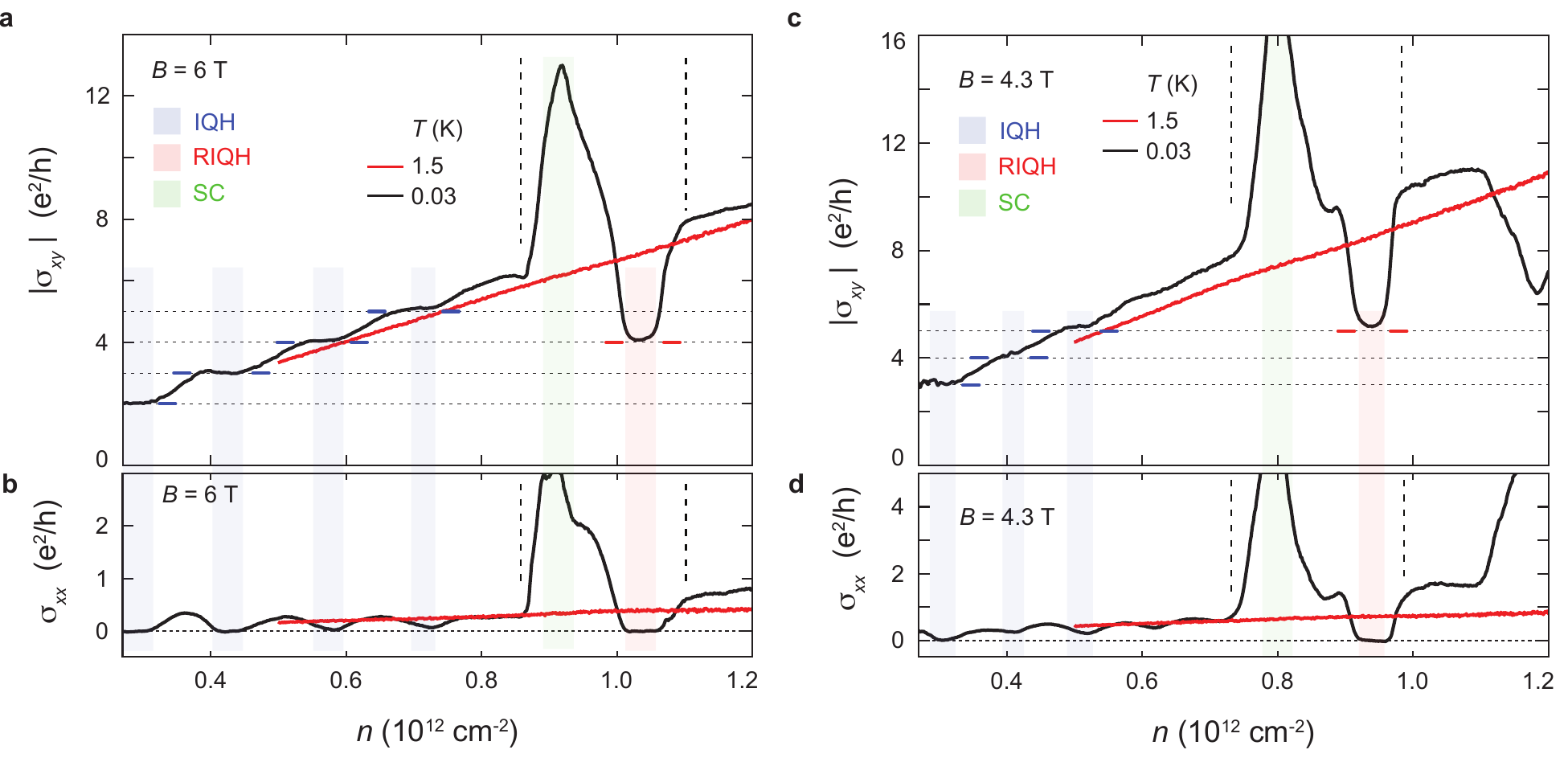}
\caption{\label{R_n_temp} \textbf{Temperature dependence.} 
(a, c) $\sigma_{xy}$ and (b, d) $\sigma_{xx}$ as a function of $n$ measured at (a--b) $B_{\perp} = 6$ T and (c--d) $B_{\perp} = 4.3$ T. Black and red traces are measured at $T = 30$ mK and $T = 1.5$ K, respectively. Transport signatures of both superconductivity and RIQH states are observed at $30$ mK, whereas both phenomena disappears at $1.5$ K.}
\end{figure*}

\begin{figure*}
\includegraphics[width=0.85\linewidth]{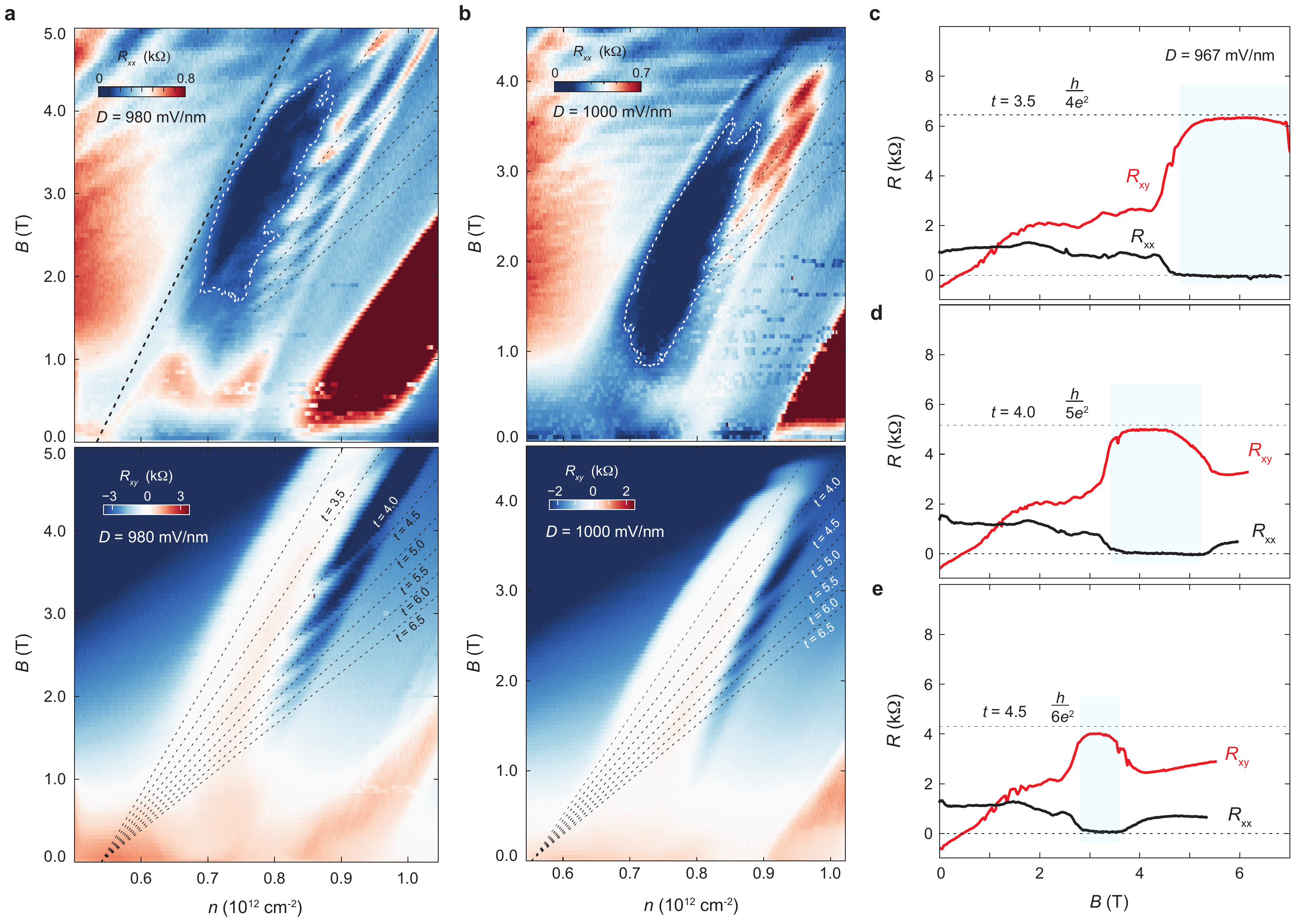}
\caption{\label{mismatch} \textbf{Fitting RIQH states with Streda slopes.} 
(a, b) $n$--$B$ maps of $R_{xx}$ (top panels) and $R_{xy}$ (bottom panels) measured at (a) $D = 980$~mV/nm and (b) $D = 1000$~mV/nm. The superconducting region is enclosed by a white dashed line in the top panels. The sequence of RIQH states is fit using Streda slopes extrapolated to the low-density side of regime~I, where the effective density of itinerant electrons is expected to vanish. The fits are determined by aligning the Streda slope with the trajectory of each RIQH state. While this procedure captures the overall trajectories, it yields a series of half-integer Streda slopes, despite the Hall plateaus being quantized at integer multiples of $h/e^2$. This mismatch can be naturally attributed to the presence of CDW order.}
\end{figure*}

We summarize below the key features of the RIQH states observed here.  

The transport signatures of the RIQH states share important similarities with those of the IQHE. At low temperature, the response exhibits vanishing longitudinal resistance and a finite Hall resistance. As shown in Fig.~\ref{fig1}g, the Hall resistance changes sign upon reversing the magnetic field direction. This behavior indicates a conductivity matrix with vanishing diagonal components and nonzero, antisymmetric off-diagonal components—characteristic of an incompressible quantum Hall state.  

There are, however, several prominent distinctions between RIQH and IQH states. Conventional IQH states typically exhibit a gradual onset as a function of temperature (Fig.~\ref{RT_QHE}a). In particular, the Hall resistance at the center of a quantized Hall plateau remains nearly temperature-independent, since its value is fixed by the charge carrier density and is unaffected by thermal fluctuations. By contrast, the RIQH states reported here all display sharply defined, temperature-dependent transitions (Figs.~\ref{RT_QHE}b--d).

We attribute this temperature-driven onset to the melting transition of the underlying CDW order. As temperature increases above the melting point, the Hall resistance $R_{xy}$ decreases substantially, consistent with a sudden increase in the effective density of itinerant electrons due to the collapse of the CDW order. At the same time, $R_{xx}$ exhibits a pronounced peak at the transition. This non-monotonic temperature dependence is a hallmark of electron solid melting, consistent with previous reports of CDW melting transitions.

Figure~\ref{R_n_temp} compares the transport response as a function of $n$, measured below and above the melting transition (black and red traces). At $T = 1.5$~K, the transport signatures associated with both the RIQH state and superconductivity vanish. Above the melting transition, $\sigma_{xy}$ increases monotonically with $n$, consistent with all charge carriers contributing as itinerant electrons once the CDW order is destroyed. We further note that $\sigma_{xy}$ measured outside regime~I remains largely unchanged between $30$~mK and $1.5$~K. This confirms that the observed temperature dependence within regime I arises from the formation of CDW order.

In the same vein, the stability of an IQH state depends directly on the out-of-plane magnetic field. The relevant energy gap of an IQH state is enhanced with increasing $B_{\perp}$, allowing the state to become more robust at higher temperatures. In contrast, the stability of the RIQH states observed here is largely insensitive to increasing $B_{\perp}$. As shown in Figs.~\ref{fig3}a--b and \ref{RT_QHE}, the melting transition of the underlying CDW is essentially $B$-independent. This behavior is consistent with the view that the CDW order is decoupled from Landau-level formation in the vicinity of regime~I, and is instead driven by strong Coulomb interactions together with the topological character of the underlying energy band.

Next, we discuss the trajectories of RIQH states across the $n$--$B$ planes. As described above, the CDW structure is continuously tunable throughout the $n$--$B$--$D$ phase space. This tunability gives rise to an apparent mismatch between the Streda slope and the quantization values of the Hall plateaus. For example, Figs.~\ref{mismatch}a--b show RIQH trajectories in the $n$--$B$ plane measured at different values of $D$. Transport responses taken along the trajectory of each RIQH state are shown in Figs.~\ref{mismatch}c--e, highlighting the apparent mismatch. In the presence of an underlying CDW order, changes in $n$ no longer directly reflect variations in the density of itinerant electrons, rendering the Streda slope ill-defined. Curiously, however, a single set of Streda slopes, as shown in Fig.~\ref{mismatch}, appears to capture all RIQH states. This raises an intriguing possibility that the ratio of localized and itinerant electrons evolves in a sysmetamic fashion across the $n-B$ plane. This outlines an interesting open question that will be the subject of future investigation.

Next, we discuss the trajectories of RIQH states across the $n$--$B$ planes. As described above, the CDW structure is continuously tunable throughout the $n$--$B$--$D$ phase space. This tunability gives rise to an apparent mismatch between the Středa slope and the quantization values of the Hall plateaus. For example, Figs.~\ref{mismatch}a--b show RIQH trajectories in the $n$--$B$ plane measured at different values of $D$. Transport responses taken along the trajectory of each RIQH state are shown in Figs.~\ref{mismatch}c--e, highlighting the apparent mismatch. In the presence of an underlying CDW order, changes in $n$ no longer directly reflect variations in the density of itinerant electrons, rendering the Středa slope ill-defined. Curiously, however, a single set of Středa slopes, as shown in Fig.~\ref{mismatch}, appears to capture all RIQH states. This raises the intriguing possibility that the ratio of localized to itinerant electrons evolves in a systematic fashion across the $n$--$B$ plane. Elucidating this organizing principle presents an open question for future investigation.

\begin{figure*}
\includegraphics[width=0.85\linewidth]{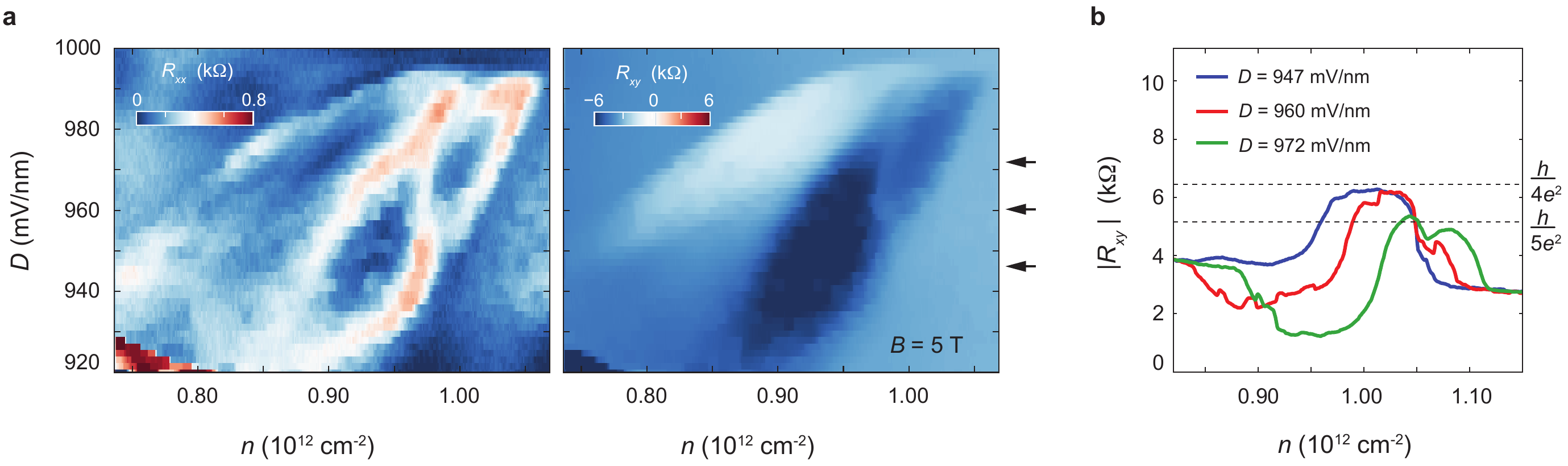}
\caption{\label{5T}{\bf{Coexisting RIQH states with different Hall plateaus.}} 
(a) $R_{xx}$ (left) and $R_{xy}$ (right) as functions of $n$ and $D$ across regime~I, measured at $B_{\perp} = 5$~T. This $n$--$D$ map captures two RIQH states simultaneously. The boundary of each state is defined by a resistance peak in $R_{xx}$, while the two states exhibit distinct quantized Hall plateaus.  
(b) $|R_{xy}|$ as a function of $n$, taken at constant $B_{\perp}$ and $D$ from the lower panel of (a). Blue and red traces, measured at $D = 947$ and $960$~mV/nm, respectively, show Hall plateaus near $h / 4e^{2}$. The green trace, taken at $D = 972$~mV/nm, exhibits a plateau near $h / 5e^{2}$.  
The re-entrant Hall plateaus evolve along trajectories in the $n$--$D$ plane, where their positions in $n$ shift systematically with $D$. This indicates that the fraction of itinerant carriers varies with displacement field, implying that the structure of the underlying CDW order changes continuously with $D$. In addition, the $h / 4e^{2}$ plateau disappears around $D \approx 970$~mV/nm, apparently replaced by the superconducting phase in the large-$D$ limit of regime~I. This points to competition between superconductivity and the re-entrant quantum Hall states.
}
\end{figure*}

\clearpage

\begin{figure*}
\includegraphics[width=0.62\linewidth]{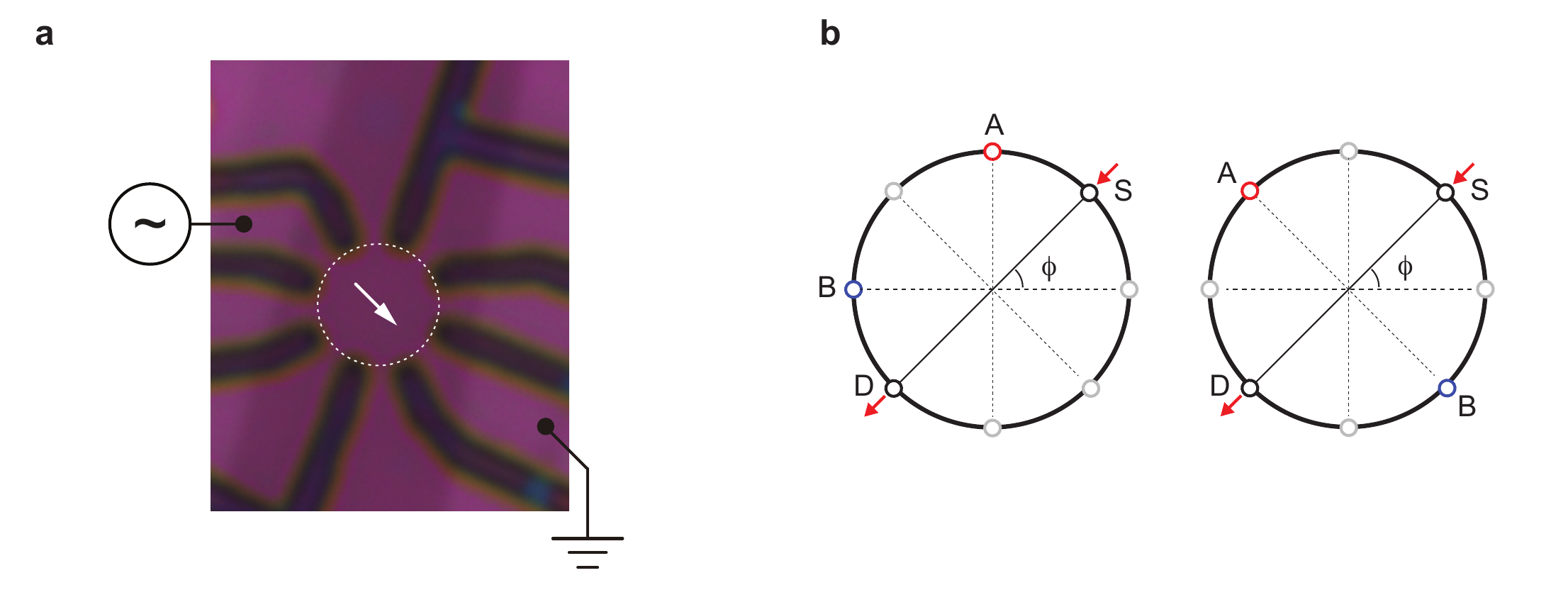}
\caption{\label{sample} \textbf{Sample geometry.} 
(a) Optical image of the R6G sample, patterned into the sunflower geometry: a disk-shaped channel with eight electrical contacts. 
(b) Schematic diagram of two measurement configurations. In the left (right) panel, voltage is measured between contacts aligned parallel (perpendicular) to the current flow direction. 
In the quantum Hall regime, these configurations yield measurements consistent with the longitudinal and transverse resistances, $R_{xx}$ and $R_{xy}$. 
In angle-resolved transport measurements, the measurement angle $\phi$ is operationally defined by the line connecting the two contacts used for current injection (source and drain). $\phi = 0$ refers to the configuration where the left and right contacts in panel (a) serve as drain and source, respectively.  
}
\end{figure*}

\begin{figure*}
\includegraphics[width=0.72\linewidth]{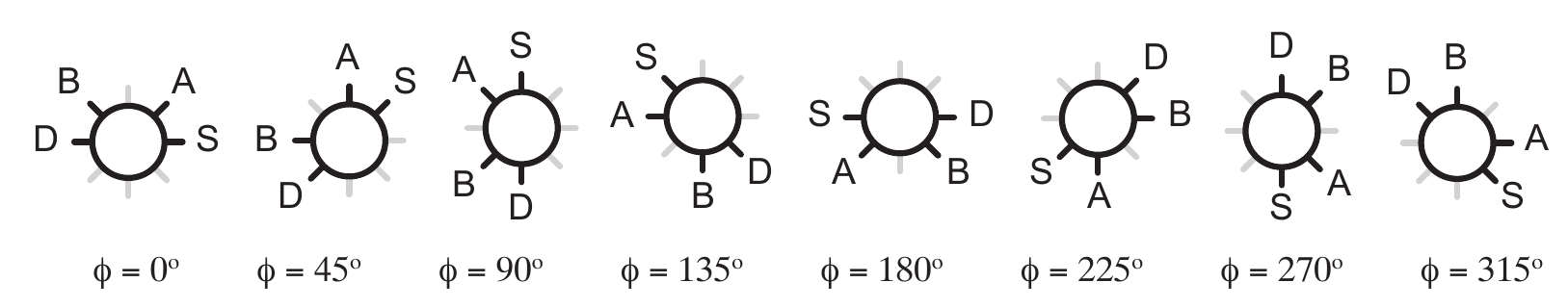}
\caption{\label{schematic} %\textbf{Measurement configurations for angle-resolved transport.} 
Measurement configurations for extracting the angular dependence of $R_{\parallel}$.  
}
\end{figure*}

\subsection{Angle-resolved transport measurement}

In this work, the R6G sample is patterned into a sunflower geometry, enabling angle-resolved transport measurements. This approach is particularly advantageous given that the key phenomena under investigation emerge from a region of phase space characterized by extreme transport anisotropy.

Figure~\ref{sample}a shows an optical image of the sample, along with typical measurement configurations used throughout this study. The sunflower geometry, in combination with angle-resolved transport techniques, has been extensively employed in previous experimental works~\cite{Zhang2024nonreciprocity,Chichinadze2024nonlinearHall,Morissette2025rhombohedral,Morissette2025stripeSC}. Here, we provide discussions of angle-resolved transport measurement. 

To perform angle resolved transport, specific measurement configurations are utilized. Fig.~\ref{sample}b shows the configurations used to probe $R_{\parallel}$ and $R_{\perp}$ at a specific angle of $\phi = 45^{\circ}$. Here, $\phi=0$ is defined by flowing current from right to left horizontally across the disk-shaped channel. Angle-resolved transport response is extracted by rotating the measurement configuration, every subsequent rotation increases $\phi$ by $45^{\circ}$. For example, Fig.~\ref{schematic} shows eight configurations used to extract the angular dependence of \Rpara. For any other configurations, the same sequence of rotations can be used to extract their corresponding angular dependence.

Angle‐resolved transport is not limited to the configurations shown in Fig.~\ref{sample}b. Given the sunflower geometry, there are 105 independent contact configurations, allowing up to 840 data points to be used in extracting the full conductivity matrix~\cite{Chichinadze2024nonlinearHall}.

\begin{figure*}
\includegraphics[width=0.66\linewidth]{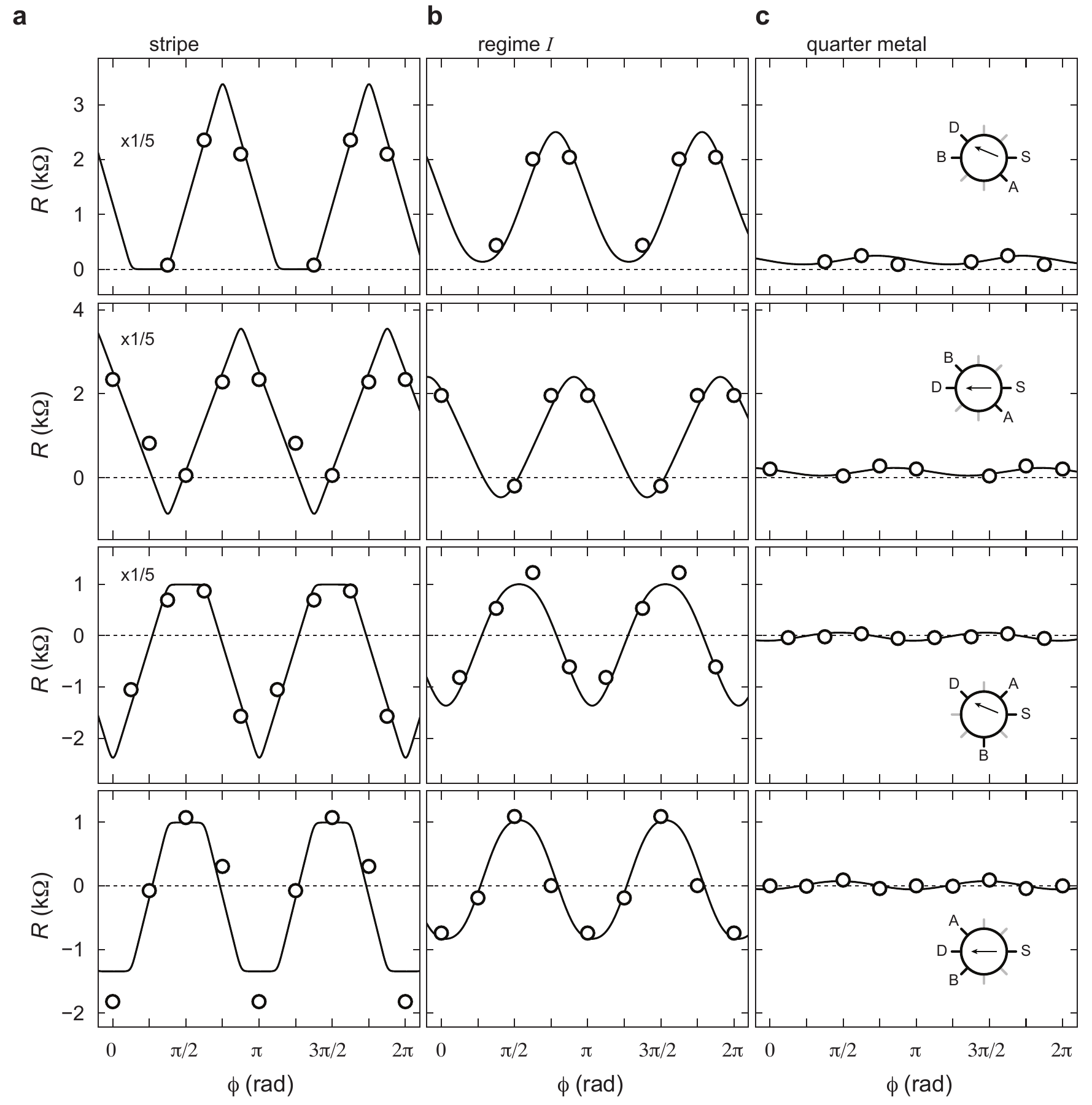}
\caption{\label{angular} \textbf{Angle-resolved transport and conductivity matrix.}  
(a--c) Angle-resolved measurements using an extended set of configurations: 
(a) inside the stripe regime at $n = 0.5 \times 10^{12}$ cm$^{-2}$ and $D = 945$ mV/nm, 
(b) inside regime~I at $n = 0.73 \times 10^{12}$ cm$^{-2}$ and $D = 980$ mV/nm, and 
(c) in the quarter-metal phase on the high-density side of the stripe regime at $n = 1.10 \times 10^{12}$ cm$^{-2}$ and $D = 980$ mV/nm. Inset in panel (c) shows the measurement configuration used for each row.
Solid black lines represent the best fits using a conductivity matrix model, yielding anisotropy ratios of 
$\sigma_{\text{max}}/\sigma_{\text{min}} = 1000$ in the stripe regime, 
$\sigma_{\text{max}}/\sigma_{\text{min}} = 4$ in regime~I, and 
$\sigma_{\text{max}}/\sigma_{\text{min}} = 1.5$ in the quarter-metal phase. 
}
\end{figure*}

Performing the full set of 840 measurements for each condition is unnecessary in most cases; instead, we typically utilize a subset of configurations. Figure~\ref{angular} shows angular responses measured with four representative configurations, with the schematic diagram of each configuration provided as the inset of panel (c). This represents a more extensive set of angle-resolved transport measurements, which is performed for three distinct phases at zero field. Figure~\ref{angular}a corresponds to the stripe order, Fig.~\ref{angular}b to regime~I, and Fig.~\ref{angular}c to the quarter-metal state on the high-density side of the stripe regime.  

Following the framework proposed in Ref.~\cite{Vafek2023anisotropy}, we perform a single global fit to all data points simultaneously, enabling the extraction of the full conductivity matrix, including the conductivities along the orthogonal principal axes. In Fig.~\ref{angular}, this global fit is shown as the black solid traces. These extensive measurements uncover the evolution of transport anisotropy, characterized by the conductivity ratio $\sigma_{\max}/\sigma_{\min}$. In the stripe phase, extreme anisotropy is observed, with $\sigma_{\max}/\sigma_{\min}$ reaching values up to $10^3$. In the presence of such extreme anisotropy, the associated error bar is relatively large, since further increasing the ratio primarily sharpens the corners of the fit without altering the overall curve.

By contrast, the angular dependence measured from regime~I exhibits much weaker anisotropy, with the conductivity ratio reduced by orders of magnitude compared to the stripe phase. Nevertheless, a finite transport anisotropy persists. In this regime, the angular response is well described by a sinusoidal oscillation with period $\pi$, in excellent agreement with previous angle-resolved transport studies where the underlying anisotropy was less extreme than in the stripe phase ~\cite{Wu2017nematic,Zhang2025angular}. The suppressed, yet non-vanishing anisotropy indicates a distorted bubble phase, with the distortion giving rise to the observed transport anisotropy.

Finally, the high-density quarter metal state is much less resistive, with the underlying anisotropy further suppressed to a ratio of $1.5$. 

Beyond extracting the conductivity matrix, this expansive dataset enables an unambiguous identification of a uniform sample, in which the transport response across many electrical contacts is governed by a single, well-defined conductivity matrix.

\begin{figure*}
\includegraphics[width=0.92\linewidth]{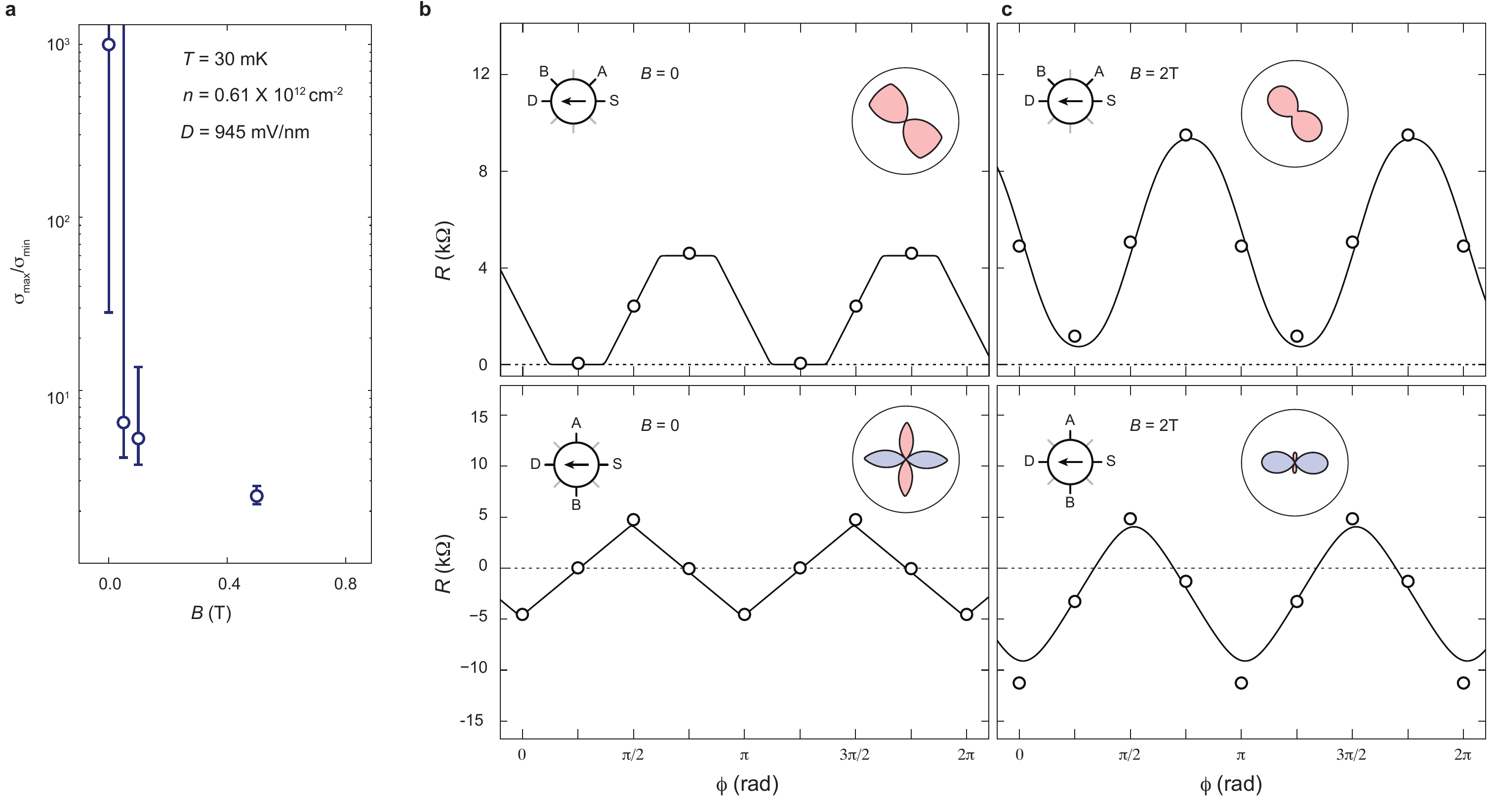}
\caption{\label{B_angle} \textbf{$B_{\perp}$ dependence of the stripe order.} 
(a) Anisotropy ratio $\sigma_{max}/\sigma_{min}$ as a function of $B_{\perp}$ measured from the regime of stripe phase. (b--c) Angle-resolved measurements of  longitudinal (top panels) and transverse responses (bottom panels), measured at (b) zero field in the presence of the stripe order and (c) $B_{\perp} = 2$ T, where anisotropy is substantially suppressed.
}
\end{figure*}

Fig.~\ref{B_angle} shows the evolution of transport anisotropy with an out-of-plane magnetic field. At zero field, the angular dependence shown in Fig.~\ref{B_angle}b reveals an extreme transport anisotropy, characterized by a large conductivity ratio of $\sigma_{\max}/\sigma_{\min} \approx 10^3$. With increasing $B_{\perp}$, the anisotropy ratio is rapidly suppressed. Fig.~\ref{B_angle}c shows the angular dependence measured at $B_{\perp} = 2$ T, which corresponds to a ratio of $\sigma_{\max}/\sigma_{\min} \approx 3$.

\subsection{Regimes I, II, and III}

\begin{figure*}
\includegraphics[width=0.8\linewidth]{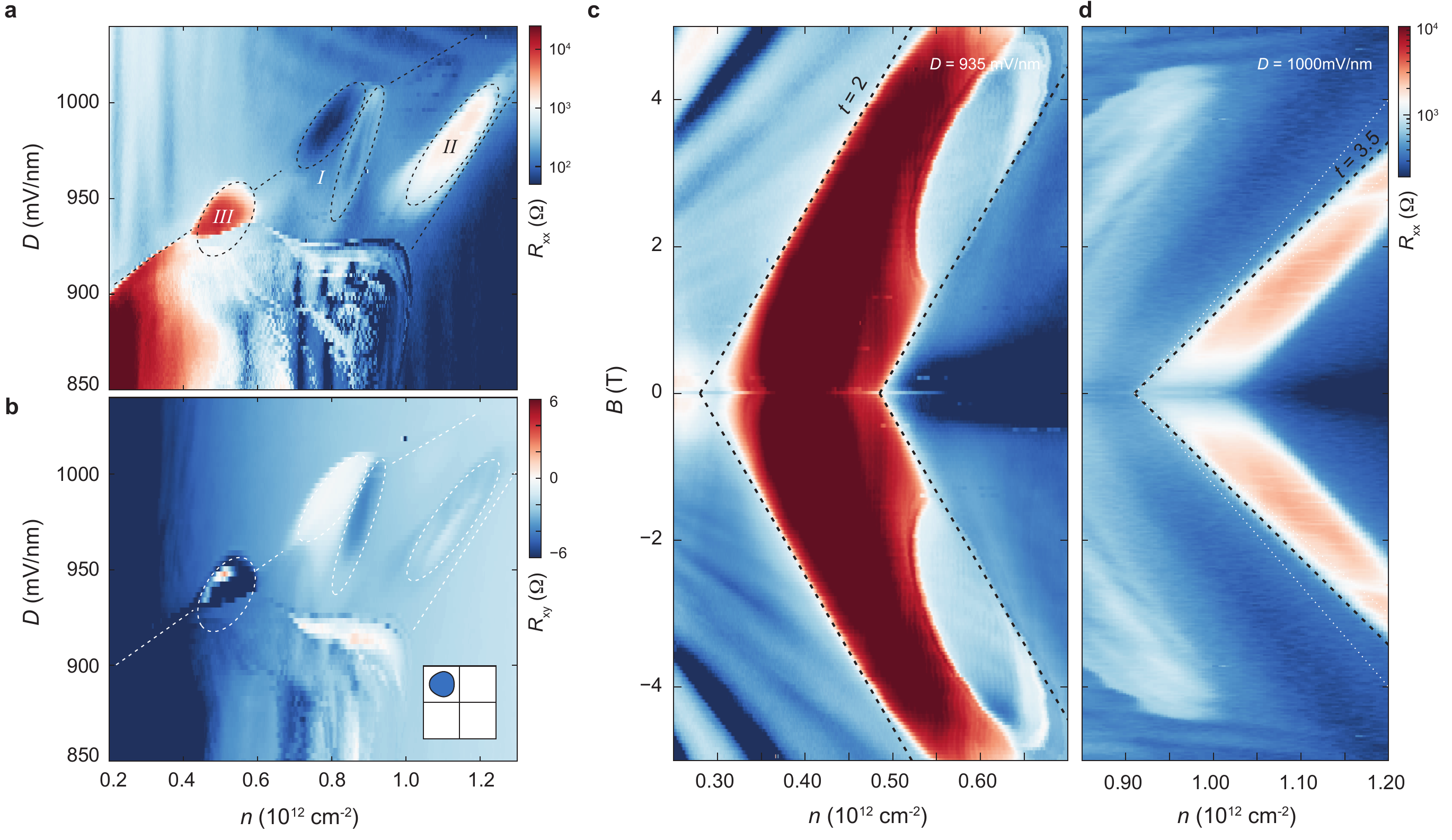}
\caption{\label{Pockets} \textbf{Regimes I, II, and III.} 
(a--b) $n$--$D$ maps of (a) $R_{xx}$ and (b) $R_{xy}$ measured at $B = 3$~T. The superconducting and RIQH states in regime~I are indicated with separate dashed circles. 
(c--d) Longitudinal resistance $R_{xx}$ plotted across the $n$--$B$ plane, highlighting the evolution of regime~(c)~III and (d)~II. In panel (d), the black dashed line marks the Streda slope $t = 3.5$, while the white dotted lines mark the integer slopes $t = 3$ and $t = 4$. The boundary of regime~II, highlighted by the black dashed line, clearly deviates from the lines with integer slopes.}
\end{figure*}

While the stripe order is suppressed by $B_{\perp}$, the metallic phases identified as regimes~I, II, and III remain clearly identifiable based on their distinct transport responses compared to the surrounding phase space. Figures~\ref{Pockets}a--b show the $n$--$D$ maps of $R_{xx}$ and $R_{xy}$ measured at $B_{\perp} = 3$~T. Here, the boundaries of regime~I are naturally defined by the superconducting phase on the low-density side and the RIQH states on the high-density side.  

Regimes~II and III exhibit somewhat different transport characteristics. In the $n$--$D$ map at $B_{\perp} = 3$~T, both are delineated by prominent resistive peaks in $R_{xx}$ (Fig.~\ref{Pockets}a). At the same time, regime~II shows only weak variations in $R_{xy}$, whereas regime~III displays a pronounced peak in Hall resistance. Transport responses measured from both regimes display sharp temperature-driven transitions, similar to that shown in Fig.~\ref{fig2}e. A plausible interpretation is that regimes~I, II, and III all originate from bubble-like CDW orders with distinct internal structures.

Figures~\ref{Pockets}c--d show the boundaries of regimes~II and III across the $n$--$B$ plane, revealing a pronounced shift in $n$ with varying $B_{\perp}$. This behavior is qualitatively similar to that observed for regime~I and is consistent with the topological character of the underlying CDW order.

\begin{figure*}
\includegraphics[width=0.7\linewidth]{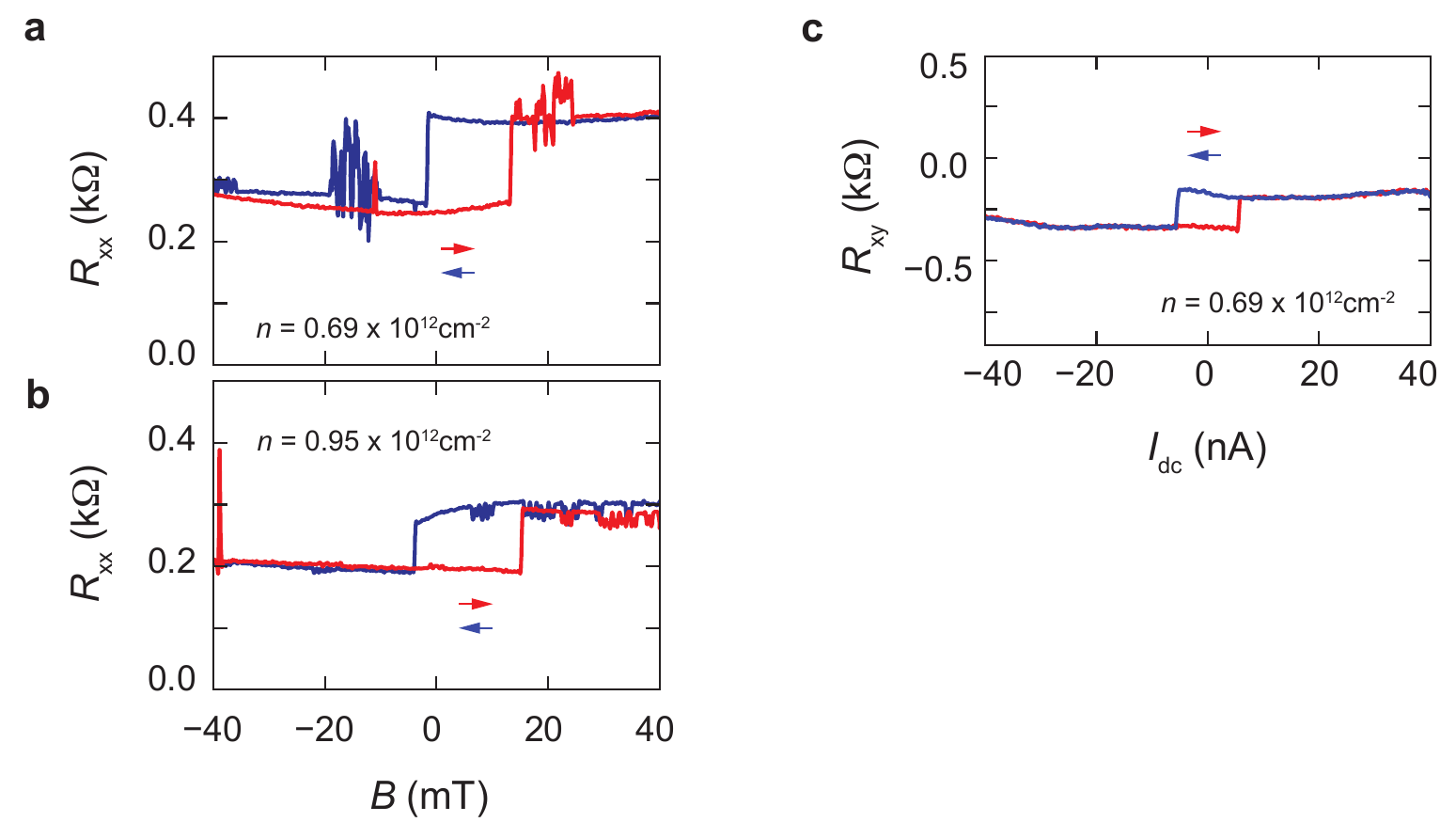}
\caption{\label{Bloop} \textbf{Hysteresis and anomalous Hall effect.} 
(a--b) Hysteresis loops in $B_{\perp}$-sweeps measured inside 
(a) regime~I and 
(b) regime~II, indicating the presence of an anomalous Hall effect. 
(c) In regime~II at $B_{\perp} = 0$, hysteretic transitions in transport response can also be induced by sweeping the d.c. current bias back and forth. 
Together, these observations are consistent with orbital ferromagnetism in regimes~I and II.
}
\end{figure*}

\begin{figure*}
\includegraphics[width=0.67\linewidth]{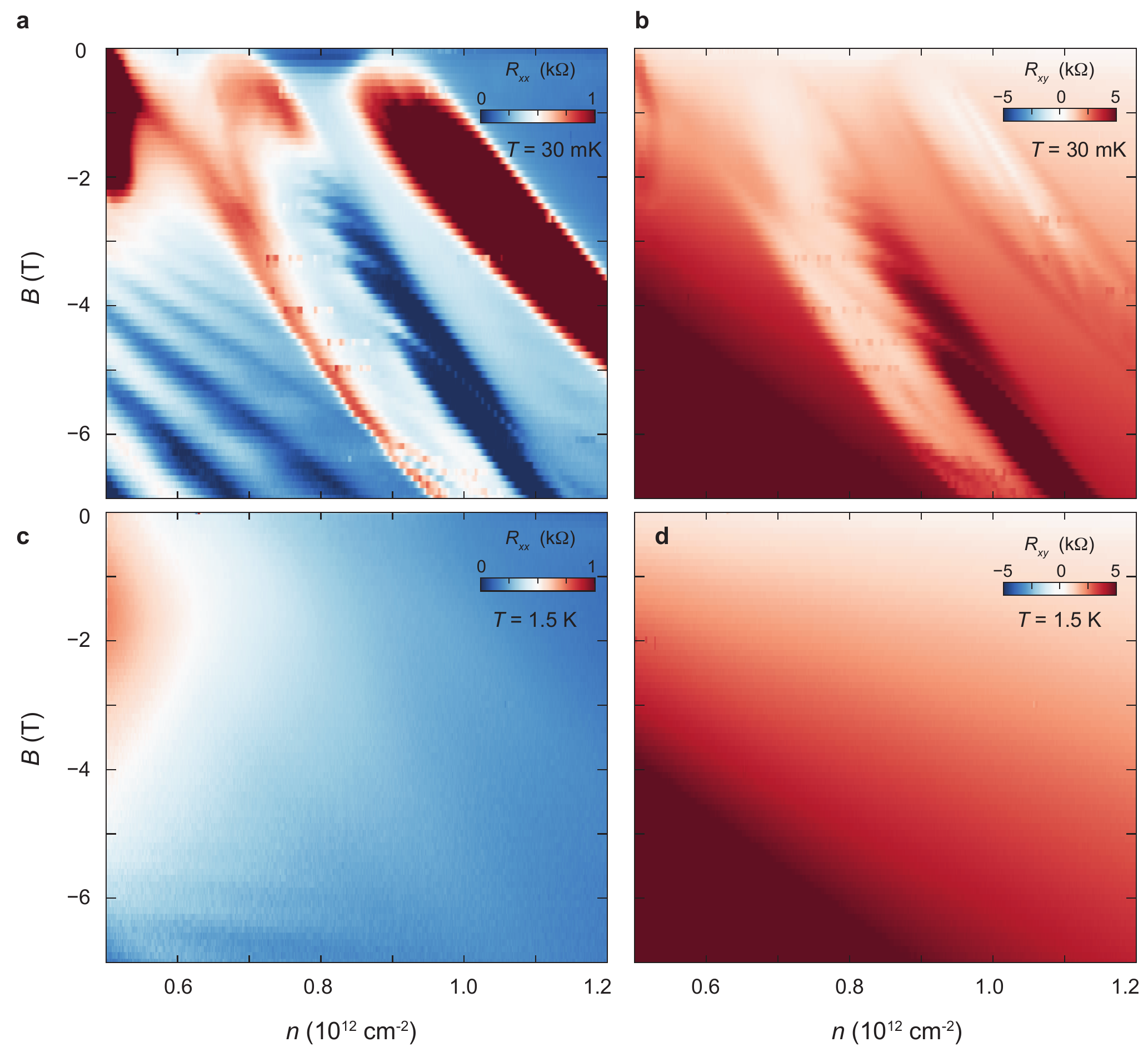}
\caption{\label{PnBT} 
\textbf{$n$--$B$ plane at 1.5~K.} 
(a--b) $n$--$B$ maps of (a) $R_{xx}$ and (b) $R_{xy}$ measured at $T = 30$~mK. 
(c--d) $n$--$B$ maps of (c) $R_{xx}$ and (d) $R_{xy}$ measured at $T = 1.5$~K. 
At $T = 1.5$~K, the boundary of regime~I, along with its associated superconducting and RIQH states, disappears.
}
\end{figure*}

The region of phase space between regimes~I and II corresponds to the stripe-ordered phase at $B_{\perp} = 0$. Notably, SdH oscillations are absent within this intermediate regime, presenting an experimental challenge for identifying the underlying electronic order. However, the presence of an anomalous Hall effect—observed in both regime I and II, as shown in Fig.~\ref{Bloop}—suggests that the system is at least partially valley polarized.

\end{document}